\documentclass[aps,prd,eqsecnum,12pt,showpacs,preprintnumbers,nofootinbib,superscriptaddress]{revtex4}
\usepackage{lscape}
\usepackage{indentfirst}
\usepackage{latexsym}
\usepackage{multirow}

\usepackage{amssymb,amsmath}
\usepackage{graphicx}
\newcommand{\bea}{\begin{eqnarray}}
\newcommand{\eea}{\end{eqnarray}}
\newcommand{\be}{\begin{equation}}
\newcommand{\ee}{\end{equation}}
\newcommand{\bes}{\begin{subequations}}
\newcommand{\ees}{\end{subequations}}
\def\lag{\langle}
\def\rag{\rangle}

\def\nn{\nonumber \\}

\begin{document}

\title{Stochastic Force Due to a Quantum Scalar Field in Minkowski Spacetime}

\author{Jason D. Bates}
\email{jdbates@staff.tku.edu.tw}
\affiliation{Department of Physics, Tamkang University, Tamsui, New Taipei City, Taiwan}

%\date{}

\begin{abstract}

\end{abstract}

\pacs{04.62+v }

\begin{abstract}

A method is presented for computing approximate expressions for the stochastic force term $\xi_{ab}$ which appears in the Einstein-Langevin equation of stochastic gravity.  Within this framework, $\xi_{ab}$ is a stochastic tensor field whose probability distribution mimics the probability distribution of the fluctuations of the quantum stress tensor operator; it is defined to be a random tensor field of zero mean whose correlation function is given by the expectation value of the symmetrized two point function of the stress energy fluctuation operator, called the noise kernel.  Approximate expressions are obtained by means of a truncated Karhunen-Loeve transform defined on a random lattice of spacetime points.  Due to the singular nature of the noise kernel, a coarse graining procedure is used to regulate divergences; as a result, the expressions obtained for $\xi_{ab}$ approximate values which might be seen by a probe measuring fluctuations in the stress energy using a sampling profile of finite width.  Two realizations of $\xi_{ab}$ in Minkowski spacetime for the conformally invariant quantum scalar field in the Minkowski vacuum state are presented.

\end{abstract}

\maketitle

\section{Introduction}

Stochastic gravity~\cite{stograLivRev} is a theory first developed in the '90s with the intent to extend semiclassical gravity to include the effects of the backreaction of quantum fluctuations of matter fields on the geometry of spacetime.  The mechanism by which this is accomplished is the introduction of a stochastic tensor field $\xi_{ab}$ which, although classical in nature, nevertheless replicates the probabilistic behavior of the quantum stress tensor fluctuation operator $\hat{t}_{ab} \equiv \hat{T}_{ab} - \lag \hat{T}_{ab} \rag$.

This fluctuation term is incorporated into the field equations for gravity via the Einstein-Langevin equation~\cite{martin99b, ELE},
\bea 
  G_{ab}(x) = 8\pi G(\langle \hat{T}_{ab}(x)\rangle+\xi_{ab}(x)) \,, \label{e-l}
\eea
wherein the stochastic tensor field $\xi_{ab}$ plays the role of the source driving the metric perturbations.  Solutions to this equation will be a stochastic distribution of geometries determined by the properties of $\xi_{ab}$.

Up to its second moment, $\xi_{ab}$ may be determined by the expectation value and two point correlation function of $\hat{t}_{ab}$,
\bes
\bea
  \lag \xi_{ab}(x) \rag_s &\equiv& \lag \hat{t}_{ab}(x) \rag = 0
\eea
\bea
  \lag \xi_{ab}(x) \xi_{c'd'}(x') \rag_s &\equiv& \lag \{ \hat{t}_{ab}(x), \hat{t}_{c'd'}(x') \} \rag \;.
\eea \label{xi_def}
\ees

If the fluctuations are assumed to follow a Gaussian probability distribution this is sufficient to uniquely determine $\xi_{ab}$~\cite{martin99b}.  Thus, the critical object for which solutions must be found is the symmetrized two-point function of the stress tensor fluctuation operator, $\lag \{\hat{t}_{ab}(x), \hat{t}_{c'd'}(x') \} \rag$, which will hereafter be referred to as the noise kernel $N_{abc'd'}(x,x')$.

Since about 2000, the major focus of investigation in stochastic gravity has been on finding ways to compute usable expressions for the noise kernel in various spacetimes~\cite{HuRou07,roura99b,PH01,martin99a,martin00,EBRAH,phillips03,roura99,pn09,ChoHu11,PH03}.  However, it is difficult to see how to extract physics directly from knowledge of the noise kernel.  This paper is a first step in extending the theory, by deriving a method to compute explicit expressions for the stochastic tensor $\xi_{ab}$ from the noise kernel itself.

Unfortunately, due to the highly singular nature of the noise kernel, without some method of regularization an explicit functional form for $\xi_{ab}$ does not exist; the stress tensor fluctuations will have infinite amplitude and be discontinuous at every point.  This behavior is not expected to be physical, but is rather an artifact of the UV behavior of the field theory.  To rectify this issue, we consider fluctuations not in the continuum limit but rather those which might be seen by some measurement apparatus of finite sampling width.  This is approximated by coarse graining through the use of smearing functions, along the lines of~\cite{FFR}.

Explicit expressions for the coarse grained fluctuations are obtained by expanding the stochastic force $\xi_{ab}$ via a Karhunen-Loeve transformation~\cite{KLT}, wherein $\xi_{ab}$ is decomposed into a series of orthogonal basis functions with random, uncorrelated coefficients of known probability distribution,
\bea
  \xi_{ab}(x) = \sum_i Z_i \phi^{(i)}_{ab}(x) \, . \label{xi-klt}
\eea
For the case of Gaussian fluctuations, each $Z_i$ will be an independent, Gaussian random variable of zero mean, and the functions $\phi^{(i)}_{ab}(x)$ are computed from the noise kernel.

Typically, a series expansion along the lines of Eq.~\eqref{xi-klt} will contain a countably infinite number of terms (one for each point in a countable, dense subset of the background spacetime), so in practice we find approximate expressions by truncating the series.  The advantage of the Karhunen-Loeve transformation is that the series so generated minimizes the mean squared error introduced by the truncation procedure.  As a general result, when $\xi_{ab}$ is assumed to be Gaussian, each of the functions $\phi^{(i)}_{ab}(x)$ may be approximated by a finite sum of functions which go like the noise kernel with one of the points fixed.

As an unexpected advantage, this coarse graining procedure provides a natural means for using fluctuations to test the validity of semiclassical gravity.  Although smearing functions introduce an artificial length scale into the theory, there is strong dependence on this length scale of the amplitude of the observed stress energy fluctuations.  Correspondingly, for lengths scales in which the amplitude of these fluctuations exceed some threshold value with high probability one would expect the semiclassical approximation to no longer hold~\cite{kuo-ford, HuRouVer04, And-Mol-Mot-1}.

This paper is organized as follows. In Sec.~\ref{klt}, a brief review is given covering the relevant details of the Karhunen-Loeve transformation.  In Sec.~\ref{snk}, a computation of the smeared noise kernel in Minkowski spacetime is presented for the conformally invariant scalar field in the Minkowski vacuum state.  In Sec.~\ref{gsef}, we derive an expression for the coarse grained stochastic force in terms of the noise kernel, and present a modified Einstein-Langevin equation which is potentially solvable.  For the case of the Minkowski vacuum, two realizations of the stochastic force are computed numerically and displayed.  Section~\ref{disc} contains a summary and discussion of the main results.  Throughout this paper, units are used such that $\hbar = c = G = k_B = 1$ and the sign convention is that of Misner, Thorne, and Wheeler~\cite{MTW}.

\section{Karhunen-Loeve Expansion}
\label{klt}

A random field $R(x)$ with zero mean and correlation function $C(x,y)$ may be expanded by means of a Karhunen-Loeve transformation~\cite{KLT} as
\bea R(x) = \sum_i Z_i r^{(i)}(x) \;, \label{klt-general} \eea
where the $Z_i$'s are centered, uncorrelated random variables and each $r^{(i)}(x)$ is a known (non-random) function of $x$.  The functions $r^{(i)}(x)$ are determined by the solutions to the eigenfunction equation
\bea \int dy \, C(x,y) r^{(i)}(y) = \lambda_i  r^{(i)}(x) \;. \label{klt-eigenf} \eea
The $Z_i$'s are determined, up to their second moment, by 
\bes
\bea \langle Z_i \rangle_s = 0 \eea
\bea \langle Z_i Z_j \rangle_s = \delta_{ij} \lambda_i \, ,\eea \label{z-moments}
\ees
where $\lag ... \rag_s$ indicates that we are taking the stochastic expectation value (as opposed to the expectation value of a quantum operator).

In general, the eigenfunction equation~\eqref{klt-eigenf} is difficult to solve exactly and there will be an infinite number of eigenfunctions.  However, it is possible to numerically approximate $R(x)$ by discretizing to a lattice of points.  Then the eigenfunction equation~\eqref{klt-eigenf} becomes a much simpler eigenvector problem
\bea \sum_{j} C(x_i,y_j) r^{(k)}(y_j) = \lambda_k  r^{(k)}(x_i) \;, \label{klt-eigenv} \eea
which may be solved using standard linear algebra techniques.  Approximate eigenfunctions may be recovered via
\bea  r^{(k)}(x) \approx \lambda_k^{-1} \sum_{i} C(x,x_i) r^{(k)}(x_i) \;. \label{klt-approx-eigenf} \eea

When the correlation function $C(x,y)$ is finite, continuous, and positive definite the series expansion in Eq.~\eqref{klt-general} converges in the mean-squared sense; that is,
\bea \lim_{k \rightarrow \infty} \lag (R(x) - R_k(x))^2 \rag_s = 0 \;, \label{klt-converge} \eea
where 
\bea R_k(x) = \sum_{i=1}^k Z_i r^{(i)}(x) \;. \eea
Furthermore, the expansion procedure presented here has the benefit that it minimizes the mean-squared error introduced due to trunctation of the series, although the accuracy of such a discretization will depend on the density of points in the lattice.

Additionally, if the random field $R(x)$ is Gaussian in nature, then each coefficient $Z_i$ is an independent Gaussian random variable and the correlation function $C(x,y)$ is sufficient to exactly specify all of the moments of each coefficient.  In this case, the summation in Eq.~\eqref{klt-general} converges exactly (not just in the mean-squared sense).  If the random field is non-Gaussian, then $C(x,y)$ is in general not sufficient to uniquely specify each $Z_i$, and the higher moments must be computed from the higher moments of $R(x)$.

Examining Eq.~\eqref{klt-approx-eigenf}, it is apparent that each approximate eigenfunction is constructed from a finite sum of functions which look like the correlation function $C(x,y)$ with one point fixed.  We will call such functions 
\bea \Psi^{(i)}(x) \equiv C(x,x_i) \label{shapef} \eea
shape functions.

By reformulating the eigenvector equation as a matrix equation,
\bea 
  \left( \begin{array}{ccc} C(x_1,x_1) & \cdots & C(x_1,x_N) 
                                         \\ \vdots & \ddots & \vdots 
                                         \\  C(x_N,x_1) & \cdots & C(x_N,x_N) \end{array} \right)
     \left( \begin{array}{c} r^{(i)}(x_1) \\ \vdots \\ r^{(i)}(x_N) \end{array} \right)
    &=& \lambda_i  \left( \begin{array}{c} r^{(i)}(x_1) \\ \vdots \\r^{(i)}(x_N) \end{array} \right) \; , \label{klt-matrix}
\eea
it is clear that each column of the correlation matrix $C$ corresponds to one shape function $\Psi^{(i)}$.  Let $A$ be the unitary matrix of eigenvectors $r^{i}$ such that
\bea A^{T} \cdot C \cdot A = \textit{diag}(\lambda) \; . \eea
Written thusly, it is clear by inspection of Eq.~\eqref{klt-approx-eigenf} that each approximate eigenfunction $r^{(k)}(x)$ is given by
\bea r^{(k)}(x) \approx \lambda_k^{-1} \sum_i A_{i,k} \Psi^{(i)}(x) \;. \eea
Finally, we may write an approximate expression for the random field in terms of the eigenvalues, eigenvectors, shape functions, and random coefficients as
\bea R(x) \approx \sum_{i,k} Z_k \lambda_k^{-1} A_{i,k} \Psi^{(i)}(x) \;. \eea
This provides a computationally inexpensive way to approximately obtain realizations for any Gaussian random field provided that the correlation function is known and satisfies the finiteness, continuity, and positive definiteness criteria.

\section{Smeared Noise Kernel}
\label{snk}

A general expression for the noise kernel for the conformally invariant scalar field was derived in~\cite{PH01} in terms of four derivatives acting on the square of the Wightman function,
%*************************************************************************
%*************************************************************************
%
%    NOISE KERNEL FUNCTIONAL FORM, GENERAL EXPRESSION
%
%*************************************************************************
%*************************************************************************
%
\begin{equation}
 N_{abc'd'}  =
 {\rm Re} \left\{  \bar K_{abc'd'}
  + g_{ab}   \bar K_{c'd'}
 + g_{c'd'} \bar K'_{ab}
 + g_{ab}g_{c'd'} \bar K \right\}
\label{general-noise-kernel}
\end{equation}
with\footnote{Note that the superscript $+$ on $G^+$ has been omitted for notational simplicity.}
\bes
\label{General-Noise}
\begin{eqnarray}
%=========================================================================
%          Display N_{abcd} terms,     GENERAL EXPRESSION, conformal
%=========================================================================
9  \bar K_{abc'd'} &=&
%
%     pf = 1 of 9
4\,\left( G{}\!\,_{;}{}_{c'}{}_{b}\,G{}\!\,_{;}{}_{d'}{}_{a} +
    G{}\!\,_{;}{}_{c'}{}_{a}\,G{}\!\,_{;}{}_{d'}{}_{b} \right)
%
%     pf = 5 of 9
+ G{}\!\,_{;}{}_{c'}{}_{d'}\,G{}\!\,_{;}{}_{a}{}_{b} +
  G\,G{}\!\,_{;}{}_{a}{}_{b}{}_{c'}{}_{d'} \cr
%
%     pf = 2,4 of 9
&& -2\,\left( G{}\!\,_{;}{}_{b}\,G{}\!\,_{;}{}_{c'}{}_{a}{}_{d'} +
    G{}\!\,_{;}{}_{a}\,G{}\!\,_{;}{}_{c'}{}_{b}{}_{d'} +
    G{}\!\,_{;}{}_{d'}\,G{}\!\,_{;}{}_{a}{}_{b}{}_{c'} +
    G{}\!\,_{;}{}_{c'}\,G{}\!\,_{;}{}_{a}{}_{b}{}_{d'} \right)  \cr
%
%     pf = 3,7 of 9
&& + 2\,\left(
G{}\!\,_{;}{}_{a}\,G{}\!\,_{;}{}_{b}\,{R{}_{c'}{}_{d'}} +
    G{}\!\,_{;}{}_{c'}\,G{}\!\,_{;}{}_{d'}\,{R{}_{a}{}_{b}} \right)  \cr
%
%     pf = 6,8 of 9
&& - \left( G{}\!\,_{;}{}_{a}{}_{b}\,{R{}_{c'}{}_{d'}} +
  G{}\!\,_{;}{}_{c'}{}_{d'}\,{R{}_{a}{}_{b}}\right) G
%
%     pf = 9 of 9
 +{\frac{1}{2}}  {R{}_{c'}{}_{d'}}\,{R{}_{a}{}_{b}} {G^2}
\end{eqnarray}
\begin{eqnarray}
%=========================================================================
%          Display N_{ab} terms,     GENERAL EXPRESSION, conformal
%=========================================================================
 36  \bar K'_{ab} &=&
%
%    pf = 1,2 of 9
8 \left(
 -  G{}\!\,_{;}{}_{p'}{}_{b}\,G{}\!\,_{;}{}^{p'}{}_{a}
 + G{}\!\,_{;}{}_{b}\,G{}\!\,_{;}{}_{p'}{}_{a}{}^{p'} +
  G{}\!\,_{;}{}_{a}\,G{}\!\,_{;}{}_{p'}{}_{b}{}^{p'}
\right)\cr &&
%
%    pf = 4,5
4 \left(
    G{}\!\,_{;}{}^{p'}\,G{}\!\,_{;}{}_{a}{}_{b}{}_{p'}
  - G{}\!\,_{;}{}_{p'}{}^{p'}\,G{}\!\,_{;}{}_{a}{}_{b} -
  G\,G{}\!\,_{;}{}_{a}{}_{b}{}_{p'}{}^{p'}
\right) \cr
%
%    pf = 3,6
&& - 2\,{R'}\,\left( 2\,G{}\!\,_{;}{}_{a}\,G{}\!\,_{;}{}_{b} -
    G\,G{}\!\,_{;}{}_{a}{}_{b} \right)  \cr
%
%    pf = 7,8
&&  -2\,\left( G{}\!\,_{;}{}_{p'}\,G{}\!\,_{;}{}^{p'} -
    2\,G\,G{}\!\,_{;}{}_{p'}{}^{p'} \right) \,{R{}_{a}{}_{b}}
%
%    pf = 9
 - {R'}\,{R{}_{a}{}_{b}} {G^2}
\end{eqnarray}
\begin{eqnarray}
%=========================================================================
%          Display N terms           GENERAL EXPRESSION, conformal
%=========================================================================
 36 \bar K &=&
2\,G{}\!\,_{;}{}_{p'}{}_{q}\,G{}\!\,_{;}{}^{p'}{}^{q}
+ 4\,\left( G{}\!\,_{;}{}_{p'}{}^{p'}\,G{}\!\,_{;}{}_{q}{}^{q} +
    G\,G{}\!\,_{;}{}_{p}{}^{p}{}_{q'}{}^{q'} \right)  \cr
&& - 4\,\left( G{}\!\,_{;}{}_{p}\,G{}\!\,_{;}{}_{q'}{}^{p}{}^{q'} +
    G{}\!\,_{;}{}^{p'}\,G{}\!\,_{;}{}_{q}{}^{q}{}_{p'} \right)  \cr
&& + R\,G{}\!\,_{;}{}_{p'}\,G{}\!\,_{;}{}^{p'} +
  {R'}\,G{}\!\,_{;}{}_{p}\,G{}\!\,^{;}{}^{p} \cr
&& - 2\,\left( R\,G{}\!\,_{;}{}_{p'}{}^{p'} +
{R'}\,G{}\!\,_{;}{}_{p}{}^{p} \right)
     G
+  {\frac{1}{2}} R\,{R'} {G^2}  \;.
\end{eqnarray}
\label{generalnoise}
\ees
Primes on indices denote tensor indices at the point $x'$ and unprimed ones denote indices at the point $x$.  $R_{ab}$ and $R_{c'\,d'}$ are the Ricci tensor evaluated at the points $x$ and $x'$, respectively; $R$ and $R'$ are the scalar curvature evaluated at $x$ and $x'$.

Evaluated in Minkowski space, the noise kernel for this field in the Minkowski vacuum may be written as~\cite{BCAH},
\bea
  N_{abc'd'}(x,x') &=& \frac{\sigma_a \sigma_b \sigma_{c'} \sigma_{d'}}{48 \pi^4 \sigma^6} + \frac{\sigma_{(a} \eta_{b)(c'} \sigma_{d')}}{24 \pi^4 \sigma^5} + \frac{4 \eta_{a(c'} \eta_{d')b} - \eta_{ab} \eta_{c'd'}}{192 \pi^4 \sigma^4} \label{nk-mink} \; .
\eea
Here $(...)$ indicates symmetrization of the indices, $\eta_{ab}$ is the Minkowski metric, and $\eta_{ac'} = \text{diag}(-1, 1, 1, 1)$ is the bivector of parallel transport in Minkowski space for Cartesian coordinates.  $\sigma$ is the Synge world function, which in Minkowski space evaluates to
\bea
  \sigma(x,x') = \frac{1}{2}[-(t-t')^2+(\vec{x}-\vec{x}')^2] \;,
\eea
and $\sigma_{a}$ indicates the covariant derivative of $\sigma$ with respect to $a$.

As can be seen by inspection of Eq.~\eqref{nk-mink}, the noise kernel diverges in the coincident limit and for null separations of points as $1/(x-x')^8$.  As a result of this divergence, no functional expression for $\xi_{ab}$ exists; each eigenvalue in Eq.~\eqref{klt-eigenf} will be infinite.  Similar to the case of forces in Brownian motion, $\xi_{ab}(x)$ will be infinite and discontinuous at every point.

In order to rectify this problem, it is necessary to find a method to regulate the noise kernel.  The method used here is coarse graining through the use of smeared operators, following the procedure outlined in~\cite{FFR}.  For our purposes, it is sufficient to smear only along the time coordinate.\footnote{Spatial-only smearing is also sufficient to regulated the divergences, as is shown in~\cite{HuRou07}.}

Using a Lorentzian smearing function
\bea
  W(t) = \frac{\alpha}{\pi (t^2 + \alpha^2)} \;, \label{smearf}
\eea
we compute the smeared Wightman function in Cartesian coordinates as
\bea
  \widetilde{G}(x,x') &\equiv& \int dt'' \, dt''' \, G(\vec{x},\vec{x}',t,t') W(t''-t) W(t'''-t') \nn
  &=& \frac{ \text{$\Delta $t}^2 - 4 \alpha^2 - r^2}{4\pi^2 [ r^4 -2r^ 2 ( \text{$\Delta $t}^2 - 4 \alpha^2) + ( \text{$\Delta $t}^2 + 4 \alpha^2)^2]} \;, \label{smearG}
\eea
where $ \text{$\Delta $t} = (t-t')$ and $r = \sqrt{(\vec{x}-\vec{x}')^2}$.  The resulting expression for the $\widetilde{N}_{0000}$ component of the smeared noise kernel is\footnote{For brevity, only the $\widetilde{N}_{0000}$ is displayed.  The expressions for the other components are similar, but depending on the component may have explicit dependence on separations in the x, y, and z coordinates.}
\bea
  \widetilde{N}_{0000}(x,x') &=& \frac{1}{12\pi^2 [ r^4 -2r^ 2 ( \text{$\Delta $t}^2 - 4 \alpha^2) + ( \text{$\Delta $t}^2 + 4 \alpha^2)^2]^6} \times \nn
                            & & \left[ \left(r^2-\text{$\Delta $t}^2\right)^6 \left(3 r^4+10 r^2 \text{$\Delta
   $t}^2+3 \text{$\Delta $t}^4\right) \right. \nn
                            & & \left. -32 \alpha ^2 \text{$\Delta $t}^2 \left(r^2-\text{$\Delta
   $t}^2\right)^4 \left(33 r^4+42 r^2 \text{$\Delta $t}^2+5
   \text{$\Delta $t}^4\right) \right. \nn
                            & & \left. -64 \alpha ^4 \left(r^2-\text{$\Delta $t}^2\right)^2 \left(9 r^8+132
   r^6 \text{$\Delta $t}^2-526 r^4 \text{$\Delta $t}^4-212 r^2
   \text{$\Delta $t}^6+21 \text{$\Delta $t}^8\right) \right. \nn
                            & & \left. -512 \alpha ^6 \left(6 r^{10}-95 r^8 \text{$\Delta $t}^2-552 r^6
   \text{$\Delta $t}^4+390 r^4 \text{$\Delta $t}^6-246 r^2
   \text{$\Delta $t}^8-15 \text{$\Delta $t}^{10}\right) \right. \nn
                            & & \left. + 512 \alpha ^8 \left(45 r^8+1300 r^6 \text{$\Delta $t}^2+1326 r^4
   \text{$\Delta $t}^4+436 r^2 \text{$\Delta $t}^6+221 \text{$\Delta
   $t}^8\right) \right. \nn
                            & & \left. + 8192 \alpha ^{10} \left(36 r^6+309 r^4 \text{$\Delta $t}^2+110 r^2
   \text{$\Delta $t}^4+57 \text{$\Delta $t}^6\right) \right. \nn
                            & & \left. + 49152 \alpha ^{12} \left(25 r^4+78 r^2 \text{$\Delta $t}^2+21
   \text{$\Delta $t}^4\right) \right. \nn
                            & & \left. + 131072 \alpha ^{14} \left(18 r^2+13 \text{$\Delta $t}^2\right) \right. \nn
                            & & \left. + 1769472 \alpha ^{16} \right] \; . \label{nk-mink-smear}
\eea
In the $\alpha = 0$ limit this reduces to
\bea
  N_{0000} &=&   \frac{3 r^4+10 r^2 \text{$\Delta $t}^2+3 \text{$\Delta $t}^4}{12\pi^2 [ r^2- \text{$\Delta $t}^2]^6} \; ,
\eea
which agrees exactly with Eq.~\eqref{nk-mink}. 

\begin{figure}
\includegraphics[width=2.5in,clip]{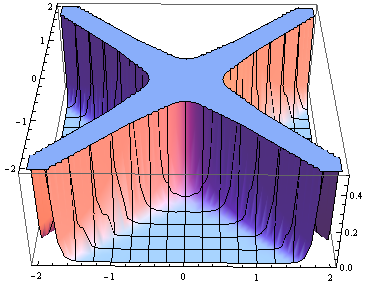}
\caption{$N_{0000}$ component of the unsmeared noise kernel in the x-t plane ($y=z=0$).}
\label{fig1}
\end{figure}

\begin{figure}
\includegraphics[width=2.5in,clip]{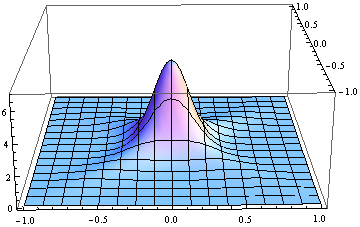}
\caption{$\widetilde{N}_{0000}$ component of the smeared noise kernel in the x-t plane ($y=z=0$) for $\alpha=0.25$.}
\label{fig2}
\end{figure}

 Figures~\ref{fig1} and~\ref{fig2} show the comparison between the $N_{0000}$ component of the unsmeared and the smeared noise kernel; in the smeared case all divergences have vanished and the function is continuous and finite everywhere.

\section{Gaussian Stress Energy Fluctuations}
\label{gsef}

The goal is to find a Karhunen-Loeve expansion for the stochastic tensor field $\xi_{ab}(x)$ of the form given in Eq.~\eqref{xi-klt} following the scheme outlined in Sec.~\ref{klt}.  In this case, the eigenfunctions $\phi^{(i)}_{ab}$'s are given by solutions of the eigenfunction equation
\bea 
  \int d^4x' \, \sqrt{-g(x')}\, N_{ab~~}^{~~c'd'}(x,x') \phi^{(i)}_{c'd'}(x') = \lambda_k\, \phi^{(i)}_{ab}(x) \, , \label{xi-eigenf}
\eea
where the only change from Eq.~\eqref{klt-eigenf} is the addition of the sum over the primed indices.  As before, if the fluctuations are assumed to have a Gaussian probability distribution, then the noise kernel is sufficient to uniquely specify $\xi_{ab}$.

Discretizing to a lattice of points gives the eigenvector equation
\bea 
  \sum_{j=0}^{N} N_{ab}^{~~c'd'}(x_i,x_{j}) \phi^{(k)}_{c'd'}(x_{j}) = \lambda_k \, \phi^{(k)}_{ab}(x_i) \, , \label{xi-eigenv}
\eea
which can be solved using standard linear algebra techniques.  In order to prevent any artificial periodic structure from appearing as a result of this discretization procedure, the points in the lattice are chosen at random from a closed volume of spacetime.  As the error in the approximation scheme is related to the density of points, it is therefore possible that regions of higher error will occur due to a locally underdense sampling of points; however, this issue can be controlled by the choice of a sufficiently large number of points.

As in Sec.~\ref{klt}, shape functions are constructed in terms of the noise kernel, giving
\bea 
  \Psi^{(j,l)}_{ab}(x) \equiv c^{(l)}_{c'd'} N_{ab}^{~~c'd'}(x,x_{j}) \, . \label{xi-shapef}
\eea
However, in contrast with Eq.~\eqref{shapef}, here it is necessary to introduce constant tensors $c^{(l)}$ to pick out individual components of the noise kernel.  For simplicity, choose each $c^{(l)}$ such that a single component is equal to one and all other components are zero.  For instance, choose $c^{(1)}_{cd} \equiv \delta_{c}^{0} \delta_{d}^{0}$.  Thus, there are 16 functions $\Psi_{ab}(x)$ for each point $x_i$ in the lattice; however, due to the symmetries present in the noise kernel only nine of these are linearly independent.

Reformulating as a matrix equation gives
\bea 
  & \left( \begin{array}{ccc} N_{ab}^{~~c'd'}(x_1,x_1) & \cdots & N_{ab}^{~~c'd'}(x_1,x_N) 
                                         \\ \vdots & \ddots & \vdots 
                                         \\  N_{ab}^{~~c'd'}(x_N,x_1) & \cdots & N_{ab}^{~~c'd'}(x_N,x_N) \end{array} \right)
     \left( \begin{array}{c} \phi^{(k)}_{c'd'}(x_1) \\ \vdots \\ \phi^{(k)}_{c'd'}(x_N) \end{array} \right) & \nn
  & = \,\,\,\, \lambda_k  \left( \begin{array}{c} \phi^{(k)}_{c'd'}(x_1) \\ \vdots \\ \phi^{(k)}_{c'd'}(x_N) \end{array} \right) & \, , \label{matrixeq}
\eea
where each $N_{ab}^{~~c'd'}(x_i,x_j)$ represents a $16 \times 16$ block of the matrix (and each $\phi^{(k)}_{c'd'}(x_i)$ is a 16 element block of the vector).  As before, each column of the above matrix for the noise kernel represents one of the shape functions $\Psi^{(j,l)}_{ab}(x)$ evaluated on the points of the lattice.

The final expression for $\xi_{ab}(x)$ becomes\footnote{The purpose of the $16j+l$ index of eigenvector matrix $A$ is to distinguish between the 16 possible shape functions associated with each point of the lattice.}
\bea 
  \xi_{ab}(x) &=& \sum_{j,k,l} Z_{k} \lambda_{k}^{-1} A_{k,(16j+l)} \Psi^{(j,l)}_{ab}(x) \;, \label{xi-klt-final}
\eea
where once again we define $A$ to be the matrix of eigenvectors of $N_{abc'd'}$.

We are now in a position to write down the smeared Einstein-Langevin equation,
\bea 
  \widetilde{G}_{ab}(x) &=& 8\pi G \left[ \langle \hat{T}_{ab}(x)\rangle+\sum_{j,k,l} Z_{k} \lambda_{k}^{-1} A_{k,(16j+l)} \Psi^{(j,l)}_{ab}(x) \right] \;. \label{smeared-EL}
\eea

Solutions to this equation will involve a stochastic distribution of geometries; however, within the linearized gravity approximation, for small fluctuations (large $\alpha$) it may be sufficient to compute the metric perturbation generated by each shape function $\Psi^{(j,k)}_{ab}$ individually.  Schematically, one may solve Eq.~\eqref{smeared-EL} for the metric perturbations induced by the stochastic force using~\cite{HuRouVer04}
\bea
  h_{ab}(x) &=& 8\pi \int d^4x' \, \sqrt{-g(x')} \, \mathcal{G}^{~~c'd'}_{ab}(x,x') \xi_{c'd'}(x') \nn
                   &=& 8\pi \sum_{j,k,l} Z_{k} \lambda_{k}^{-1} A_{k,(16j+l)} \int d^4x' \, \sqrt{-g(x')} \, \mathcal{G}^{~~c'd'}_{ab}(x,x') \Psi^{(j,l)}_{c'd'}(x') \; ,
\eea
where $\mathcal{G}^{~~c'd'}_{ab}(x,x')$ is the semiclassical retarded Green function for the operator acting on $h_{ab}$.  Thus, the set of shape functions $\Psi^{(j,l)}$ may be used to generate a basis for the induced metric pertubations.  For maximally symmetric spacetimes such as Minkowski space, one would need to compute backreactions for only nine unique functions, due to the translation symmetry of the shape functions.

\begin{figure}
\includegraphics[width=2.5in,clip]{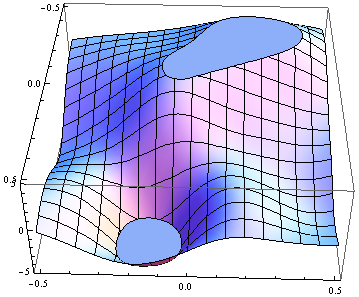}
\caption{One realization of $\xi_{ab}$ along the x-t plane ($y=z=0$) with $\alpha=0.25$ for a random lattice of 50 points chosen within the hypercube bounded by $[-0.5,0.5]$.}
\label{fig3}
\end{figure}

\begin{figure}
\includegraphics[width=2.5in,clip]{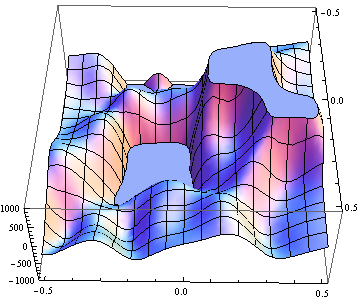}
\caption{One realization of $\xi_{ab}$ along the x-t plane ($y=z=0$) with $\alpha=0.1$ for a random lattice of 50 points chosen within the hypercube bounded by $[-0.5,0.5]$.}
\label{fig4}
\end{figure}

Even without solving the backreaction problem, it is possible to discern some of the behavior of the fluctuations by examining realizations of the stochastic force $\xi_{ab}$.  The most immediately apparent of these is the strong dependence of both the amplitude and the frequency of the fluctuations on the smearing parameter $\alpha$.  As shown in Figs.~\ref{fig3} and~\ref{fig4}, as $\alpha$ decreases the amplitude and frequency of the fluctuations increases, until in the $\alpha = 0$ limit both the amplitude and frequency of the fluctuations are infinite.  This is consistent with forces in the case of Brownian motion, where in the short distance limit the forces are infinitely strong and discontinuous everywhere.

As a result of this dependence of the amplitude and frequency on the smearing parameter $\alpha$, the smeared noise kernel provides a natural length scale for use in evaluating the validity of the semiclassical approximation; since the semiclassical approximate is suspected to not be valid when fluctuations of the stress energy tensor are large, then it is unlikely to be valid on length scales smaller than some $\alpha_0$ chosen such that when $\alpha \ge \alpha_0$ the fluctuations remain below a given threshold value with high probability.

A second feature evident from Figs.~\ref{fig3} and~\ref{fig4} is that fluctuations tend to propagate along null vectors, as would be expected from the fact that the noise kernel is large for points near null separation; however, the length scale of this propagation is dependent on $\alpha$, with small $\alpha$ leading to shorter propagation lengths due to the higher frequency of interfering fluctuations.

\section{Discussion}
\label{disc}

In this paper, we have derived a method for generating approximate, coarse grained expressions for the stochastic force term $\xi_{ab}$ which appears in the Einstein-Langevin equation of stochastic gravity.  This method makes use of a Karhunen-Loeve transform to express $\xi_{ab}$ in terms of decorrelation random coefficients and known non-random functions of the spacetime points.  In order to render this method amenable to numerical computation, the Karhunen-Loeve transform was discretized to a random lattice of points.

Coarse graining proved necessary due to the highly singular nature of the noise kernel and the resulting pathological UV behavior of the fluctuations.  While this behavior is not expected to be physical, it does demand that some method be chosen to regulate the noise kernel.  The method chosen here was smearing along the time direction by use of a Lorentzian smearing function with an associated length parameter $\alpha$; this approximates the stress energy fluctuations which might be seen by a probe with a sampling profile which matches that of the smearing function.

The expressions for the stochastic force generated by this procedure are composed of finite summations of shape functions which look like the coarse grained noise kernel with one point fixed multiplied by random coefficients.  When the fluctuations are assumed to be Gaussian in nature, each of these coefficients is also Gaussian.

As a result of the dependence of the amplitude of the stochastic force upon the smearing parameter $\alpha$, this method provides a natural means by which the validity of semiclassical gravity may be tested.  Since semiclassical gravity is expected to fail when fluctuations are large, then it can be assume to be valid only for length scales greater than some $\alpha_0$ chosen such that when $\alpha \ge \alpha_0$ the fluctuations remain below a given threshold value with high probability.

Several realizations of $\xi_{ab}$ have been computed in Minkowski spacetime for the conformally invariant scalar field in the Minkowski vacuum state, and two such realizations have been displayed.  By investigating these realizations it is possible to discern some of the behavior of the stress energy fluctuations.  For the case of the conformally invariant scalar field, one sees that fluctuations tend to propogate along null vectors, as one would expected from the fact that correlations in the stress energy for that field are large for nearly-null separations of points.  However, this propogation length appears to be tied to the smearing parameter $\alpha$ due to the higher frequency of interfering fluctuations.

Finally, the method presented here provides a pathway to finding explicit solutions of the coarse-grained Einstein-Langevin equation.  Such solutions would be a stochastic distribution of geometries whose properties are determined by the shape functions $\Psi^{(j,k)}_{ab}$ and the random coefficients $Z_{i}$.  The author hopes to revisit this problem and compute such solutions in future work.

\section*{Acknowledgements}
The author would like to thank Paul Anderson and Hing Tong Cho for helpful conversations.  This work was supported in part by the National Science Council of the Republic of China under the grants NSC 99-2112-M-032-003-MY3, NSC 101-2811-M-032-005, and by the National Center for Theoretical Sciences (NCTS).


\begin{thebibliography}{100}

\bibitem{stograLivRev} See e.g. B. L. Hu and E. Verdaguer, Liv. Rev. Rel. {\bf 11}, 3 (2008) and references contained therein.

\bibitem{martin99b} R. Mart\'\i n and E. Verdaguer, Phys. Rev. D {\bf 60}, 084008 (1999).

\bibitem{ELE} E. Calzetta and B. L. Hu,    Phys. Rev. D {\bf 49}, 6636 (1994);
B. L. Hu and A. Matacz,    Phys. Rev. D {\bf 51}, 1577 (1995);  B. L.
Hu and S. Sinha,       Phys.  Rev. D {\bf 51}, 1587 (1995);
  A. Campos and E. Verdaguer,   Phys. Rev. D {\bf 53}, 1927 (1996);
  F. C. Lombardo and F. D. Mazzitelli,  Phys. Rev. D {\bf 55}, 3889 (1997).
  A. Campos and B. L. Hu, Phys. Rev. D {\bf 58} (1998) 125021

\bibitem{HuRou07} B. L. Hu  and Albert Roura, Phys. Rev. D {\bf 76}, 124018 (2007).

\bibitem{roura99b} A. Roura and E. Verdaguer, Int. J. Theor. Phys. {\bf 38}, 3123 (1999).

\bibitem{PH01} N. G.  Phillips and  B. L. Hu, Phys. Rev. D {\bf 63}, 104001 (2001).

\bibitem{martin99a} R. Mart\'\i n and E. Verdaguer, Phys. Lett. B {\bf 113}, 465 (1999).

\bibitem{martin00} R. Mart\'\i n and E. Verdaguer, Phys. Rev. D {\bf 61}, 124024 (2000).

\bibitem{EBRAH} A. Eftekharzadeh, J. D. Bates, A. Roura, P. R. Anderson, and B. L. Hu, Phys. Rev. D{\bf 85}, 044037 (2012).

\bibitem{phillips03} N. G. Phillips and B. L. Hu, Phys.Rev. D {\bf 67}, 104002 (2003).

\bibitem{roura99} A. Roura and E. Verdaguer, Int. J. Theor. Phys. {\bf 38}, 3123 (1999).

\bibitem{pn09} G. P\`{e}rez-Nadal, A. Roura, and E. Verdaguer, JCAP {\bf 05} 036, (2010).

\bibitem{ChoHu11} H. T. Cho and B. L. Hu, Phys. Rev. D84, 044032 (2011);
      J. Physics (Conf. Ser.) 330, 012002 (2011)  [arXiv:1105.5302]

\bibitem{PH03} N. G. Phillips  and B. L. Hu, Phys. Rev. D {\bf 67}, 104002  (2003).

\bibitem{FFR} C. J. Fewster, L. H. Ford, and T. A. Roman,  Phys. Rev. D {\bf 81}, 121901 (2010).

\bibitem{KLT} See e.g. R. Ghanem and P. D. Spanos, {\em Stochastic finite element: a spectral approach,} (Springer, New York, 1991).


\bibitem{kuo-ford}
C.-I. Kuo and L.H. Ford,
Phys. Rev. D {\bf 47}, 4510 (1993).

\bibitem{HuRouVer04} B. L. Hu, A. Roura, and E. Verdaguer,
 Phys. Rev. D {\bf 70}, 044002 (2004).

\bibitem{And-Mol-Mot-1} P. R. Anderson, C. Molina-Paris, and E. Mottola, Phys. Rev. D {\bf 67}, 024026 (2003).

\bibitem{MTW} C. W. Misner, K. S. Thorne, and J. A. Wheeler, {\it Gravitation} (Freeman, San Francisco, 1973).

\bibitem{BCAH} J. D. Bates, H. T. Cho, P. R. Anderson, and B. L. Hu, arXiv:1301.2501 (2013).




\end{thebibliography}
\end{document}